\documentclass[onecolumn,preprint,showpacs,superscriptaddress,nofootinbib,amsfonts,amssymb,amsmath]{revtex4-1}

\usepackage[T1]{fontenc}
\usepackage[utf8]{inputenc}
\usepackage{graphicx}
\usepackage{eucal}
\usepackage{mathtools}
\usepackage{amsmath}
\usepackage{amssymb}
\usepackage{xcolor}
\usepackage{soul}
\usepackage{caption}
\usepackage{subcaption}
\usepackage[colorlinks=true, pdfstartview=FitV, linkcolor=blue, citecolor=blue, urlcolor=magenta, breaklinks=true]{hyperref}
\usepackage{orcidlink}

\begin{document}

\title{Black hole solutions surrounded by anisotropic fluid in $f(\mathbb{T},\CMcal{T})$ gravity}

\author{Franciele M. da Silva \orcidlink{0000-0003-2568-2901}} 
\email{franmdasilva@gmail.com}

\affiliation{Departamento de F\'isica, CFM - Universidade Federal de Santa Catarina; \\ C.P. 476, CEP 88.040-900, Florian\'opolis, SC, Brazil.}

\author{Luis C. N. Santos \orcidlink{0000-0002-6129-1820}}
\email{luis.santos@ufsc.br}

\affiliation{Departamento de Física, CCEN--Universidade Federal da Paraíba; C.P. 5008, CEP  58.051-970, João Pessoa, PB, Brazil}

\affiliation{Departamento de F\'isica, CFM - Universidade Federal de Santa Catarina; \\ C.P. 476, CEP 88.040-900, Florian\'opolis, SC, Brazil.}

\author{V. B. Bezerra \orcidlink{0000-0001-7893-0265}}
\email{valdir@fisica.ufpb.br}
\affiliation{Departamento de Física, CCEN--Universidade Federal da Paraíba; C.P. 5008, CEP  58.051-970, João Pessoa, PB, Brazil}

\begin{abstract}
In this work, we investigate some extensions of the Kiselev black hole solutions in the context of $f(\mathbb{T},\CMcal{T})$ gravity. By mapping the components of the Kiselev energy–momentum tensor into the anisotropic energy–momentum tensor and assuming a particular form of $f(\mathbb{T},\CMcal{T})$, we obtain exact solutions for the field equations in this theory that carries dependence on the coupling constant and on the parameter of the equation of state of the fluid. We show that in this scenario of modified gravity some new structure is added to the geometry of spacetime as compared to  the Kiselev black hole. We analyse the energy conditions, mass, horizons and  the Hawking temperature considering particular values for the parameter of the equation of state.
\end{abstract}

\maketitle

\section{Introduction} 

Although the Schwarzschild black hole is one of the most important solutions obtained in General Relativity (GR), it does not take into account the possibility of the existence of an environment around the black hole. Recent black hole observations obtained by the Event Horizon Telescope (ETH) collaboration reveal a complex structure of matter around the event horizon associated with the massive black hole in the center of our galaxy \cite{akiyama2022first}. On the other hand, due to the acceleration of the universe, it is expected that there is some type of energy permeating spacetime that could explain the current observations, the so--called dark energy. Considering the gravitational field equations, there are many ways to take into account the existence of an environment around a black hole, depending on its nature. Accretion disks \cite{disc1,disc2} and rings \cite{disc3} as well as exotic matter \cite{kiselev1,kiselev2,fstring, kiselev3} around black holes can be considered by using an energy--momentum tensor associated with a perfect or anisotropic fluid. 

On the other hand, the more traditional way of describing GR is by the curvature depiction of gravity through the metric and curvature tensors. In this way, the usual approach to construct modified theories of gravity is to extend the Einstein--Hilbert action in GR \cite{retro28,Harko:2011kv} or to modify the Einstein field equations \cite{rastall1972generalization,retro27} and then test the modifications \cite{da2023rapidly,santos2019electrostatic}.
Alternatively, GR can also be described in terms of ``Teleparallel Equivalent of General Relativity'' (TEGR) by means of tetrad and torsion fields \cite{hayashi1979new, aldrovandi2012teleparallel,bahamonde_teleparallel_2023}. This formulation allows us to develop GR in both the Weitzenb{\"o}ck and Riemannian geometries, which makes this approach more general \cite{maluf2013teleparallel}. Analogously to the framework of GR, it is also possible to propose modified teleparallel theories of gravity \cite{hayashi1979new,harko2014f,bahamonde2017new,vilhena2023neutron}. In this work, we study the $f(\mathbb{T},\CMcal{T})$ gravity which is an extension of TEGR proposed in \cite{harko2014f}. In this theory, the gravitational action depends on a general non--minimal coupling between the torsion scalar $\mathbb{T}$ and the trace of the matter energy--momentum tensor $\CMcal{T}=g^{\mu\nu}T_{\mu\nu}$. 

When we solve Einstein equation for a fluid, we need information about the connection between the energy density and the pressure, that is, the equation of state of the matter we are considering. In this context, it was proposed a type of relation connecting energy density and pressure \cite{kiselev}, in which components of the energy--momentum tensor are associated with an exotic fluid. By taking the isotropic average over the angles, the solution of Einstein field equations with this fluid leads to the Kiselev black hole in GR. As a consequence of this approach, by choosing the parameter of the equation of state particular solutions of field equations can be obtained. For example, an  specific value of the parameter of the equation of state reproduce, in the cosmological context, the accelerating pattern \cite{kiselev}. Indeed, Kiselev black holes have been studied in the context of shadow of black holes \cite{shadow1,shadow2,shadow3,shadow4}, thermodynamics of black holes \cite{termo1,termo2,termo3,termo4,termo5,termo6,termo7,termo8}, quasinormal modes \cite{quasi1,quasi2,quasi3,quasi4,quasi5,quasi6,quasi7,quasi8,quasi9} and in modified theories of gravity \cite{heydarzade2017black,sakti_kerrnewmannutkiselev_2020,morais_thermodynamics_2022,santos2023kiselev,gogoi2023joule,ghosh_rotating_2024}. In this contribution, we obtain solutions of field equations in the context of $f(\mathbb{T},\CMcal{T})$ gravity considering the presence of a Kiselev fluid, in a similar way as discussed in \cite{kiselev,heydarzade2017black,morais_thermodynamics_2022,santos2023kiselev}. In previous works with extensions of TEGR, physical systems with spherical symmetry such as neutron stars were studied \cite{pace_quark_2017,ghosh_gravastars_2020,saleem_anisotropic_2022,salako_study_2020,ditta_anisotropic_2022,mota2023neutron}. In \cite{mota2023neutron}, the authors of this paper together with collaborators studied the existence of neutron stars with realistic equations of state in the same type of $f(\mathbb{T},\CMcal{T})$ gravity as considered here. In that work we have calculated physical properties such as mass and radius and compared it with recent experimental data, showing that $f(\mathbb{T},\CMcal{T})$ gravity can lead to improved results when compared to GR.

The paper is organized in the following way: In Section \ref{ftt}, we obtain the equations for black holes in an environment with an anisotropic fluid in $f(\mathbb{T},\CMcal{T})$ gravity. In Section \ref{solve}, we obtain a new general analytical solution for the field equations that represents a black hole surrounded by an anisotropic fluid of the Kiselev type. In Section \ref{energy}, we verify if this solution satisfies the energy conditions. In Section \ref{mht}, we apply our new solution to calculate some physical properties of the black holes. In Section \ref{special}, we analyze the consequences of the current results to some special cases of the fluids which surrounds the black holes and, lastly, in Section \ref{conclusions}, we present our final considerations and future work perspectives.

\section{Black holes surrounded by anisotropic fluid in $f(\mathbb{T},\CMcal{T})$ gravity} \label{ftt}

The modified theory of gravity that we consider in this paper takes into account the torsion of spacetime in addition to a coupling with matter represented by the trace of the energy--momentum tensor. The action that describes this scenario is written as \cite{harko2014f} 
\begin{equation}
\mathbb{S}= \int d^{4}x~~e \left[\frac{\mathbb{T}+f(\mathbb{T},\CMcal{T})}{16\pi}+\mathcal{L}_{m}\right],
\label{eq1}
\end{equation}
where $e=\text{det}(e^A_{\;\;\mu})=\sqrt{-g}$ is the determinant of the tetrads, $\mathbb{T}$ is the torsion scalar and $\CMcal{T}=g^{\mu\nu}T_{\mu\nu}$ is the trace of the energy--momentum tensor $T_{\mu\nu}$, which can be obtained from the Lagrangian for the matter distribution ${\mathcal{L}}_m$. Here we choose ${\mathcal{L}}_m = -\mathcal{P}$, where $\mathcal{P}=\frac{1}{3}\left(p_r+2p_t\right)$ 
 \cite{lm}. 

Let us assume that the function $f(\mathbb{T},\CMcal{T})$ is given by
\begin{equation}
    f\left(\mathbb{T},\CMcal{T}\right)=\lambda \mathbb{T} \CMcal{T},
\label{eq2}
\end{equation}
where $\lambda$ can be interpreted as a coupling constant of the geometry with matter fields \cite{harko2014f,salako2020study}. The coupling between the trace of the energy--momentum tensor and torsion represents an additional term to the action associated with the coupling between gravity and torsion. It is evident that when $\lambda=0$, this coupling disappears. Otherwise, if the torsion is zero, we obtain the Einstein--Hilbert action.  
The energy--momentum tensor associated with Kiselev black holes is defined in such a way that
\begin{align}
    T^{t}_{\:\:\:t}= T^{r}_{\:\:\:r}&=\rho(r), \label{e8}\\
   T^{\theta}_{\:\:\:\theta}= T^{\phi}_{\:\:\:\phi}&=-\frac{1}{2}\rho (3\omega+1),
   \label{e9}
\end{align}
where $\omega$ is the parameter of the equation of state. Equations (\ref{e8}) and (\ref{e9}) can be connected to the general anisotropic fluid expression \cite{santos00} 
\begin{equation}
    T_{\mu\nu}=-p_{t}(r)g_{\mu\nu}+(p_{t}(r)+\rho(r))U_{\mu}U_{\nu}+(p_{r}(r)-p_{t}(r))N_{\mu}N_{\nu}
    \label{eq4},
\end{equation}
where $\rho(r)$, $p_{r}(r)$ and $p_{t}(r)$ are the energy density, the radial pressure and the tangential pressure of the fluid, respectively. Taking into account black holes surrounded by an anisotropic fluid, as described previously by Eqs. (\ref{e8}), (\ref{e9}) and (\ref{eq4}), the equations of state that relates these components is written as \cite{kiselev}
\begin{align}
    p_r (r) &= - \rho(r), \\
     p_t (r) &= \frac{1}{2} (3 \omega +1) \rho(r)
    \label{eqx}
\end{align}
At this point, we can consider the variation of the action given by Eq. (\ref{eq1}), which gives us the following field equation
\begin{equation}
    G_{\mu\nu}=8\pi \left[ \left( 1+\frac{\lambda}{8\pi} \right) T_{\mu\nu} + \frac{\lambda}{8\pi} g_{\mu\nu} \left( \frac{p_r(r)+2 p_t(r)}{3} -\frac{T}{2} \right) \right]
    \label{eq7}
\end{equation}
where $G_{\mu\nu}$ is the Einstein tensor. Equation (\ref{eq7}) can be solved by assuming that the line element is spherically symmetric. In the next sections, we study in details the solutions of the field equations in the $f(\mathbb{T},\CMcal{T})$ gravity. 

\section{Solving the field equations} \label{solve}

A general line element associated with a spherically symmetric spacetime is given by  
\begin{equation}
    ds^2=B(r)dt^2-A(r)dr^2-r^2(d\theta^2+\sin\theta^2d\phi^2),
    \label{eq3}
\end{equation}
where $A(r)$ and $B(r)$ are radial functions. Thus, the four velocity  $U^{\mu}$ and the radial unit vector $N^{\mu}$ from Eq. (\ref{eq4}) are given as
 \begin{equation}
        U^{\mu}=\left( \frac{1}{\sqrt{B(r)}},0,0,0\right)
    \label{eq5},
 \end{equation}
 \begin{equation}
        N^{\mu}=\left( 0,\frac{1}{\sqrt{A(r)}},0,0\right)
    \label{eq6},
 \end{equation}
and satisfy the conditions $U_{\nu}U^{\nu}=1$, $N_{\nu}N^{\nu}=-1$ and $U_{\nu}N^{\nu}=0$. Using the line element (\ref{eq3}) in the Eq. (\ref{eq7}) we obtain the following system of differential equations: 
\begin{equation}
    \frac{A' }{r A ^2}-\frac{1}{r^2 A }+\frac{1}{r^2}= \left( 8 \pi +\frac{1}{2} \lambda (5 \omega +1) \right)\rho ,
    \label{eq10}
\end{equation}
\begin{equation}
    -\frac{B' }{r A  B }-\frac{1}{r^2 A }+\frac{1}{r^2}= \left( 8 \pi +\frac{1}{2} \lambda (5 \omega +1) \right)\rho
    \label{eq11}
\end{equation}
and
\begin{equation}
 \frac{B'' }{A  B }-\frac{B'^2}{2 A  B ^2}+\frac{B' }{r A  B }-\frac{A'  B' }{2 A ^2 B }-\frac{A' }{r A ^2}= ( 8 \pi ( 1 + 3 \omega ) + 2 \lambda (1 - \omega )) \rho.
 \label{eq12}
\end{equation}
From Eqs. (\ref{eq10}) and (\ref{eq11}) we can see that functions  $A(r)$ and $B(r)$ are related as
\begin{equation}
    A(r) = \frac{1}{B(r)}.
    \label{eq13}
\end{equation}
By substituting this result in Eqs. (\ref{eq11}) and (\ref{eq12}), and combining the resulting expressions, we get the following differential equation 
\begin{equation}
    \frac{1}{r}\left( \frac{d}{dr} (r B) -1 \right) = -\frac{8 \pi  (1+3 \omega)+4 \lambda  (1-\omega )}{16 \pi+\lambda  (1+5 \omega ) } \frac{d }{dr} \left( \frac{d}{dr} (r B) -1 \right),
    \label{eq14}
\end{equation}
which can be integrated by direct methods to give us the solution for the function $B(r)$, associated with the line element (\ref{eq3}). This way, we obtain the general solution for a black hole surrounded by an anisotropic fluid of the Kiselev type in $f(\mathbb{T},\CMcal{T})$ gravity, which can be written as
\begin{equation}
    B(r) = 1 + \frac{c_1}{r}+Z r^{-\frac{8 \pi  (1+3 \omega)+4 \lambda  (1-\omega )}{16 \pi+\lambda  (1+5 \omega ) }},
    \label{eq15}
\end{equation}
where $Z$ and $c_1$ are constants of integration. This solution can be used in the field equations in order to get an expression for energy density, which is given by
\begin{equation}
    \rho(r) = \frac{6 Z (16 \pi  \omega+\lambda  (1-3 \omega ) ) r^{-\frac{6 (8 \pi +\lambda ) (1+\omega)}{16 \pi +\lambda  (1+5 \omega ) }}}{(16 \pi +\lambda  (1+5 \omega ) )^2}.
    \label{eq16}
\end{equation}
As we can see, the explicit form of the energy density is determined directly from the field equations. As a consequence, the explicit expressions for the radial and tangential pressures are obtained using the equation of state for the anisotropic fluid. Notice that if $c_1= -2M$, the solution (\ref{eq16}) reduces to usual Schwarzschild metric in the absence of fluid and modification in the gravity ($\lambda = 0$).

\section{Energy conditions} \label{energy}

\begin{figure}[t]
     \centering
     \begin{subfigure}{0.49\textwidth}
         \centering
         \includegraphics[width=\textwidth]{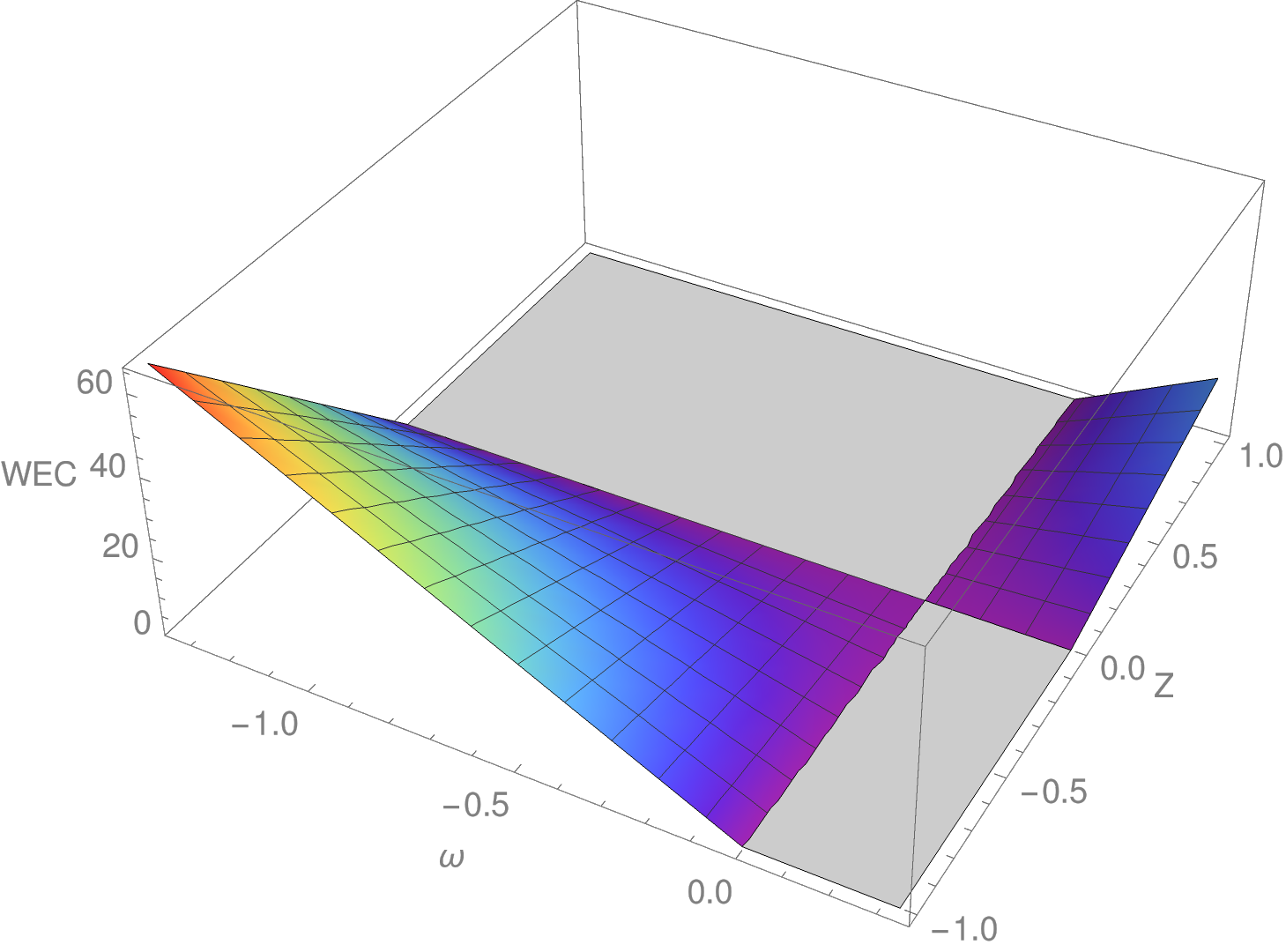}
     \end{subfigure}
     \begin{subfigure}{0.49\textwidth}
         \centering
         \includegraphics[width=\textwidth]{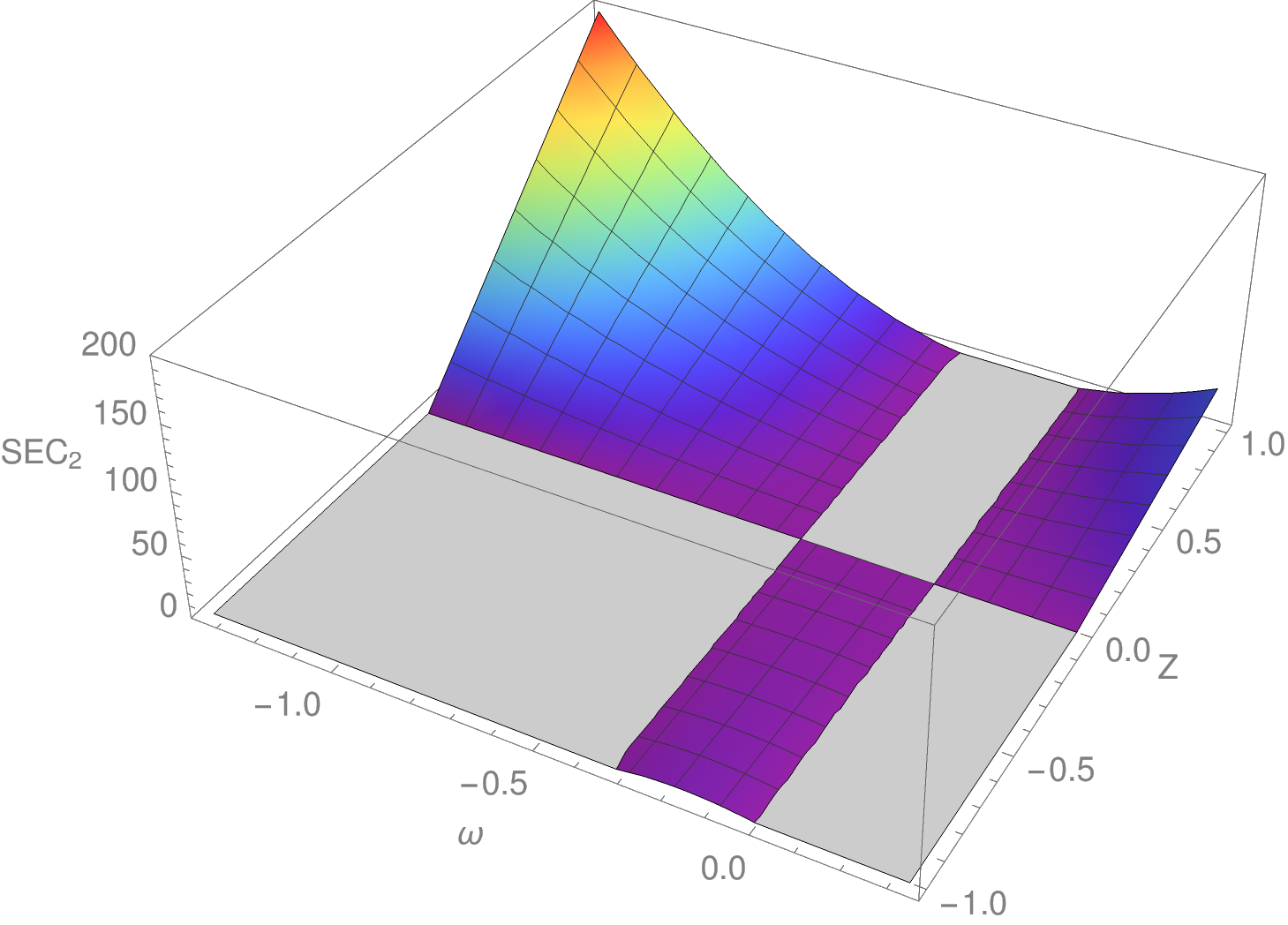}
     \end{subfigure}
        \caption{In this figure, we show the regions were the WEC is satisfied (left panel), and the regions were the SEC$_2$ is satisfied (right panel). The figures are plotted in terms of the parameters $\omega$ and $Z$, and assuming $\lambda=0.2$.}
         \label{fig1}
\end{figure}
In order to study the characteristics of the matter associated with the obtained solution, it is useful to determine the conditions under which the energy density is positive, that is, if the weak energy conditions (WEC) are satisfied. We also want to determine the requisites for the strong energy conditions (SEC) to be satisfied, \textit{i.e.}, for gravity to be an attractive force \cite{camenzind2007compact}.

The WEC is given by
\begin{equation}
    \rho  \geq 0,
    \label{eq17a}
\end{equation}
and, for an anisotropic fluid, the SEC is given by three inequalities, which we are going to call SEC$_1$, SEC$_2$ and SEC$_3$, respectively
\begin{equation}
  \rho + p_t \geq 0 \quad \quad \rho + p_r + 2 p_t \geq 0, \quad \quad \rho + p_r \geq 0,
  \label{eq17b}
\end{equation}
where, the SEC$_3$ is trivial, since $\rho = - p_r$. After replacing the expressions for the energy density and the pressures in Eqs. (\ref{eq17a}) and (\ref{eq17b}), and making some simplifications, we find two inequalities that must be obeyed so that the WEC and SEC are satisfied simultaneously. These are as follows
\begin{equation}
    Z (16 \pi  \omega+\lambda  (1-3 \omega ) ) \geq 0
    \label{eq18}
\end{equation}
and
\begin{equation}
        Z (1+3 \omega ) (16 \pi  \omega+\lambda  (1-3 \omega )) \geq 0.
        \label{eq19}
\end{equation}
We can reduce Eqs. (\ref{eq18}) and (\ref{eq19}) to the following cases:
\begin{itemize}
    \item If $Z=0$, the energy conditions are satisfied for:
\begin{equation}
    -\frac{4}{3} \leq \omega < -\frac{1}{3}.
    \label{eq20a}
\end{equation}
    \item If $Z \leq 0$, the energy conditions are satisfied for:
\begin{equation}
    -\frac{1}{3} \leq \omega <  \frac{\lambda}{3 \lambda - 16 \pi}.
    \label{eq20b}
\end{equation}
    \item If $Z \geq 0$, the energy conditions are satisfied for:
\begin{equation}
    \frac{\lambda}{3 \lambda - 16 \pi} < \omega \leq \frac{1}{3}.
    \label{eq20c}
\end{equation}
    \item In general, the energy conditions are also satisfied if:
    \begin{equation}
         \omega =  \frac{\lambda}{3 \lambda - 16 \pi}.
    \label{eq20d}
    \end{equation}  
\end{itemize}

In Fig. \ref{fig1}, the WEC (left panel) and the SEC$_2$ (right panel) are shown as a function of the constant of integration $Z$ and of the parameter $\omega$, for an specific choice of the coupling constant  $\lambda$ of the $f(\mathbb{T},\CMcal{T})$ theory. The gray regions are the ones where the energy conditions are not satisfied, and the colorful regions are the ones were the WEC and SEC$_2$ are obeyed.

\section{Mass, horizons and temperature} \label{mht}

In this section we apply our new solution, Eq. (\ref{eq15}), to calculate some physical properties of black holes.
We can determine the radius of the event horizon associated with the solution for the black hole by considering the region where the function $B(r)$ vanishes. Denoting $R_h$ as the radius of the horizon, it must satisfy the equation $B(R_h)=0$.  Thus, in terms of $R_h$ the black hole mass can be written as 
\begin{equation}
    M = \frac{1}{2} \left(R_{h}+Z R_{h}^{-\frac{3 (16 \pi  \omega+\lambda  (1-3 \omega ) )}{16 \pi+\lambda  (1+5 \omega ) }}\right).
    \label{eq21}
\end{equation}
Equation (\ref{eq21}) connects the mass of the black hole with the horizon. Considering Hawking radiation, a direct way to obtain the expression for this quantity is using the surface gravity of the black hole in the form 
\begin{equation}
    \kappa=\frac{1}{2}\frac{dB(r)}{dr}\Big|_{r=R_h},
\end{equation}
whose result is
\begin{equation}
    \kappa = \frac{R_{h}^{-\frac{6 (8 \pi+\lambda ) (1+\omega )}{16 \pi+\lambda  (1+5 \omega )  }} \left((16 \pi+\lambda  (1+5 \omega )  ) R_{h}^{\frac{\lambda
    (5+\omega)+16 \pi  (2+3 \omega )}{16 \pi+\lambda  (1+5 \omega ) }}-3 Z (16 \pi  \omega+\lambda  (1-3 \omega ) )R_h\right)}{2 (16 \pi+\lambda  (1+5 \omega )  )}.
    \label{eq22}
\end{equation}
As we can see, surface gravity  depends on the parameter of the equation of state $\omega$ and on the coupling constant of the $f(\mathbb{T},\CMcal{T})$ gravity, and thus, differ from the usual result in the context of GR. This implies that Hawking temperature $T=\hbar \kappa/(2\pi)$ has a similar structure and can be written in the form
\begin{equation}
    T = \frac{\hbar }{4 \pi  R_h}-\frac{3 Z \hbar  (16 \pi  \omega+\lambda  (1-3 \omega ) ) R_h^{-\frac{\lambda (5+\omega )+16 \pi  (2+3 \omega)}{16 \pi+\lambda  (1+5 \omega )   }}}{4 \pi  (16 \pi+\lambda  (1+5 \omega )  )}.
    \label{eq23}
\end{equation}
This temperature has correction terms due to the modified gravity as well as extra terms associated with the nature of the anisotropic fluid. In this way, it is necessary to separately analyze the influence of modified gravity and of the fluid for a given value of the horizon radius. In the next subsections we consider some important values for the parameters in the expressions obtained in this section.

\section{Special cases} \label{special}

In the next subsections, we analyze the effect of the parameter $\lambda$ of $f(\mathbb{T},\CMcal{T})$ gravity on the energy density, Eq. (\ref{eq16}), and the temperature, Eq. (\ref{eq23}) for some special cases corresponding to different values of the parameter $\omega$. In principle, the parameter $\omega$ of the equation of state is free to take any value. Here we analyze some particular values which are commonly chosen for its important physical consequences. 

\subsection{Black hole surrounded by a radiation field}

\begin{figure}[t]
     \centering
     \begin{subfigure}{0.49\textwidth}
         \centering
         \includegraphics[width=\textwidth]{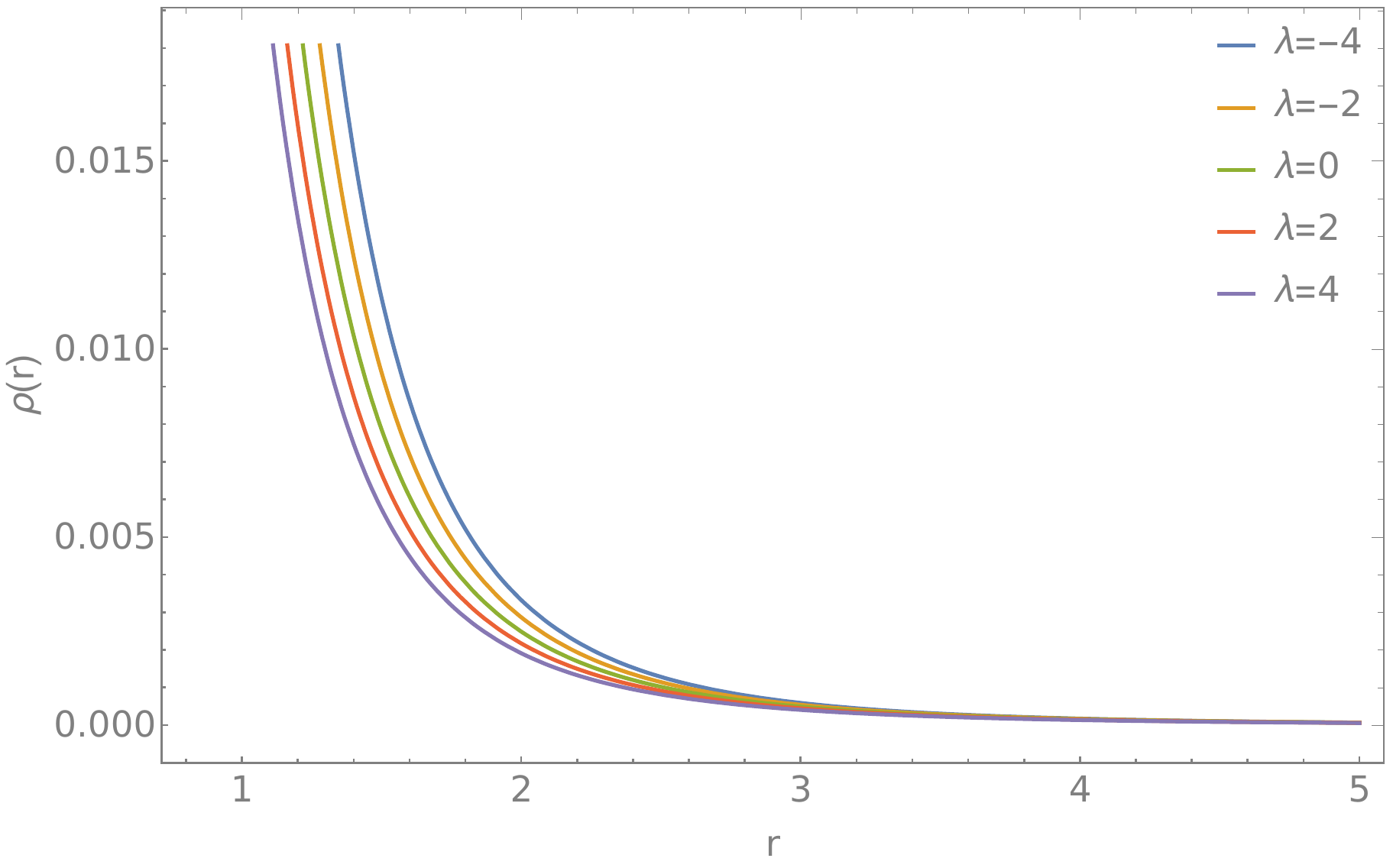}
     \end{subfigure}
     \begin{subfigure}{0.49\textwidth}
         \centering
         \includegraphics[width=\textwidth]{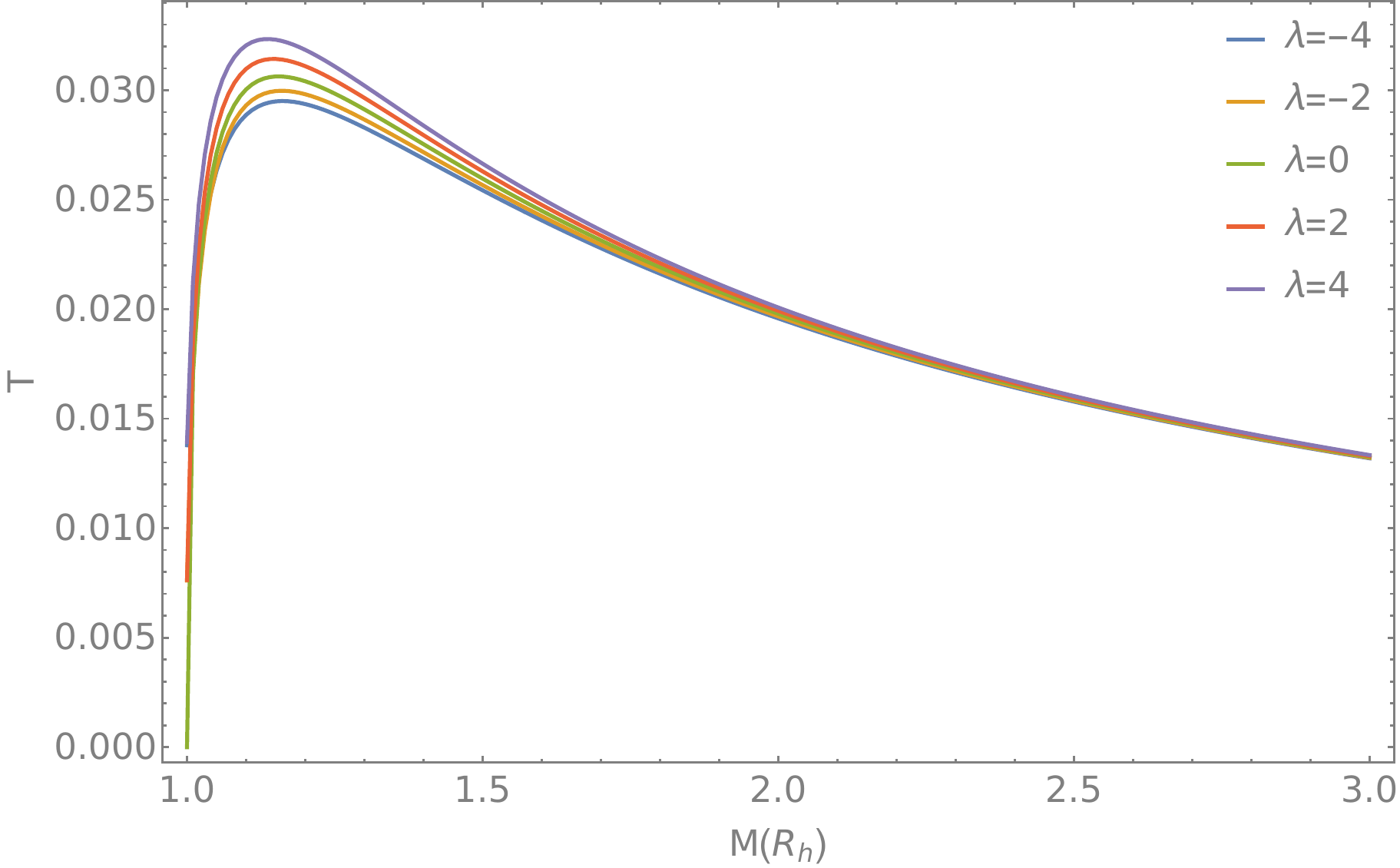}
     \end{subfigure}
        \caption{In this figure we show the energy density $\rho$ as a function of the radial coordinate $r$ (left panel), and the temperature $T$ as a function of the mass of the black hole $M(R_h)$ (right panel). These plots are for the case $\omega=1/3$ and five different values of the parameter $\lambda$, and assuming $Z=\hbar=1$. }
         \label{fig2}
\end{figure}

We start considering $\omega=1/3$ and, in which case the metric function $B(r)$ is given by
\begin{equation}
    B(r) = 1 - \frac{2 M}{r}+Z r^{-\frac{12 \pi + \lambda }{6 \pi + \lambda }}.
    \label{eq24}
\end{equation}
From Eq. (\ref{eq24}) we can observe that $B(r)$ has a dependence on the coupling constant $\lambda$, and the third term is due to the scenario in which  the $f(\mathbb{T},\CMcal{T})$ is considered. The energy density in the radiation field case is given by
\begin{equation}
    \rho(r) = \frac{9 \pi  Z r^{-\frac{3 (8 \pi + \lambda )}{6 \pi + \lambda }}}{2 (6 \pi + \lambda )^2},
    \label{eq25}
\end{equation}
so that, it depends on the radial coordinate and on $\lambda$. In the left panel of Fig. \ref{fig2} we show $\rho(r)$ given by Eq. (\ref{eq25}), from which we can conclude that the energy density diverges for $r \rightarrow 0$ and goes to zero asymptotically. The energy density decreases with the increasing of the coupling constant. As to the Hawking temperature, we find the result
\begin{equation}
    T = \frac{\hbar }{4 \pi  R_h}-\frac{3 Z \hbar  R_h^{-\frac{2 (9 \pi +\lambda )}{6 \pi + \lambda}}}{2 (6 \pi +\lambda)}.
    \label{eq26}
\end{equation}
From Eq. (\ref{eq26}) we see that $T$ changes with both $R_h$ and $\lambda$ and, in the right panel of Fig. \ref{fig2} we show the temperature as a function of the mass of the black hole. For each value of $M$ we use Eq. (\ref{eq24}) to find the horizon radius. For the cases with two horizons, if we calculate $T$ for the inner horizon we will obtain a negative temperature, in the same way as occurs with the Reissner-Nordstr\"om black hole \cite{tiandho2017implication}, which is a special case of our solution (for $\omega=0$ and $\lambda=0$). The negative temperatures found in black holes are still an open problem in the literature \cite{hayward2009local,bo2010negative,cvetivc2018killing,mcmaken2023hawking}, while some authors interpret that these negative temperatures could be associated with a negative ``pressure'' that stops the black hole from further collapse \cite{norte2024negative}, others propose modifications in the method to calculate the temperature of the black holes \cite{park2007thermodynamics,park2008can}. Analysing the right panel of Fig. \ref{fig2}, can conclude that the temperature goes to zero when $M(R_h) \rightarrow 1$, then it increases with the increasing of $M(R_h)$ until it reaches a maximum and then starts to decrease again. For a fixed mass of the black hole, the temperature increases with the increasing of $\lambda$. It is interesting to notice that for $M(R_h)  \lesssim 1$ the solutions with radiation field do not present horizons, so that we were not able to calculate $T$ for this cases.

\subsection{Black hole surrounded by a dust field}

\begin{figure}[b]
     \centering
     \begin{subfigure}{0.49\textwidth}
         \centering
         \includegraphics[width=\textwidth]{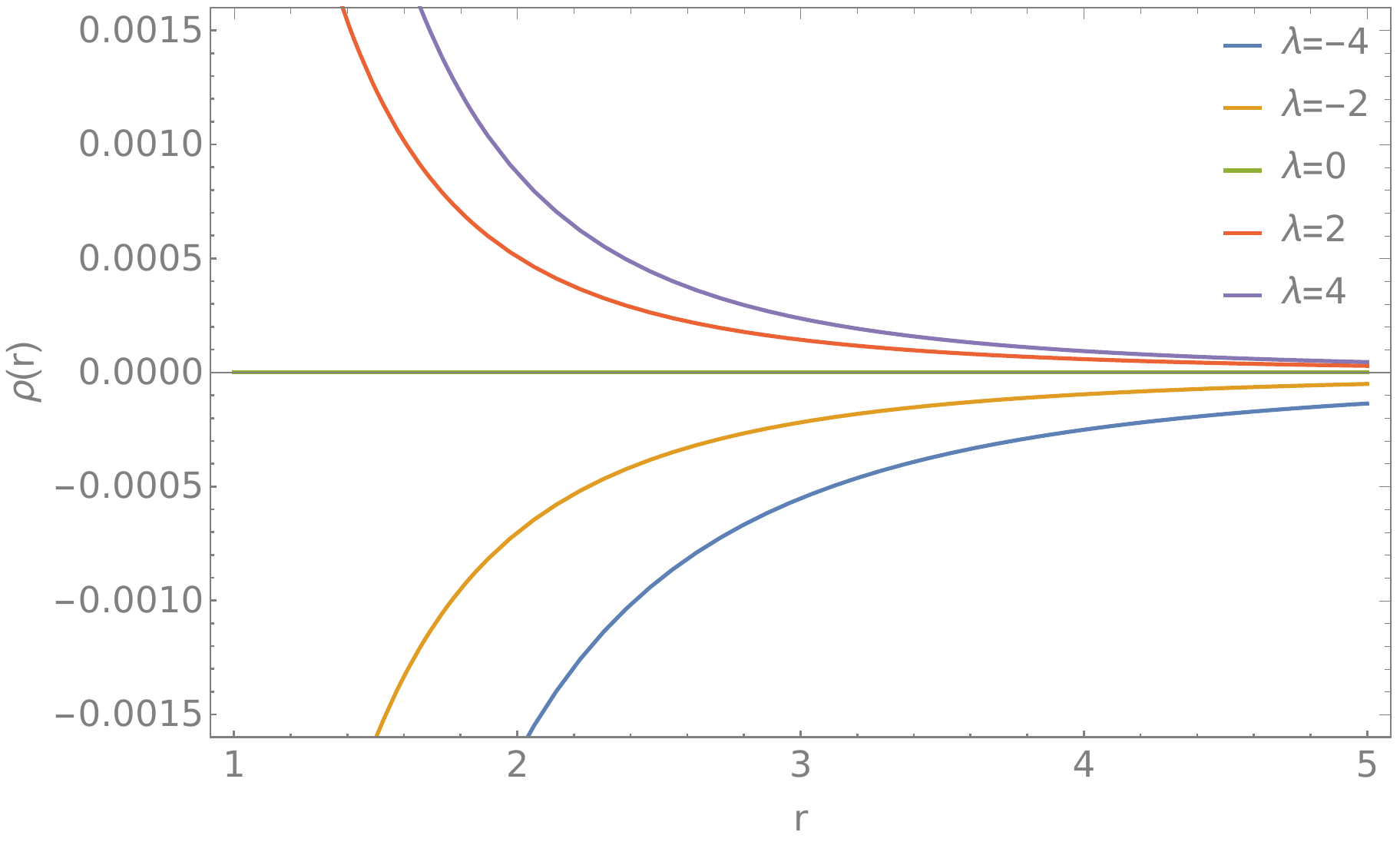}
     \end{subfigure}
     \begin{subfigure}{0.49\textwidth}
         \centering
         \includegraphics[width=\textwidth]{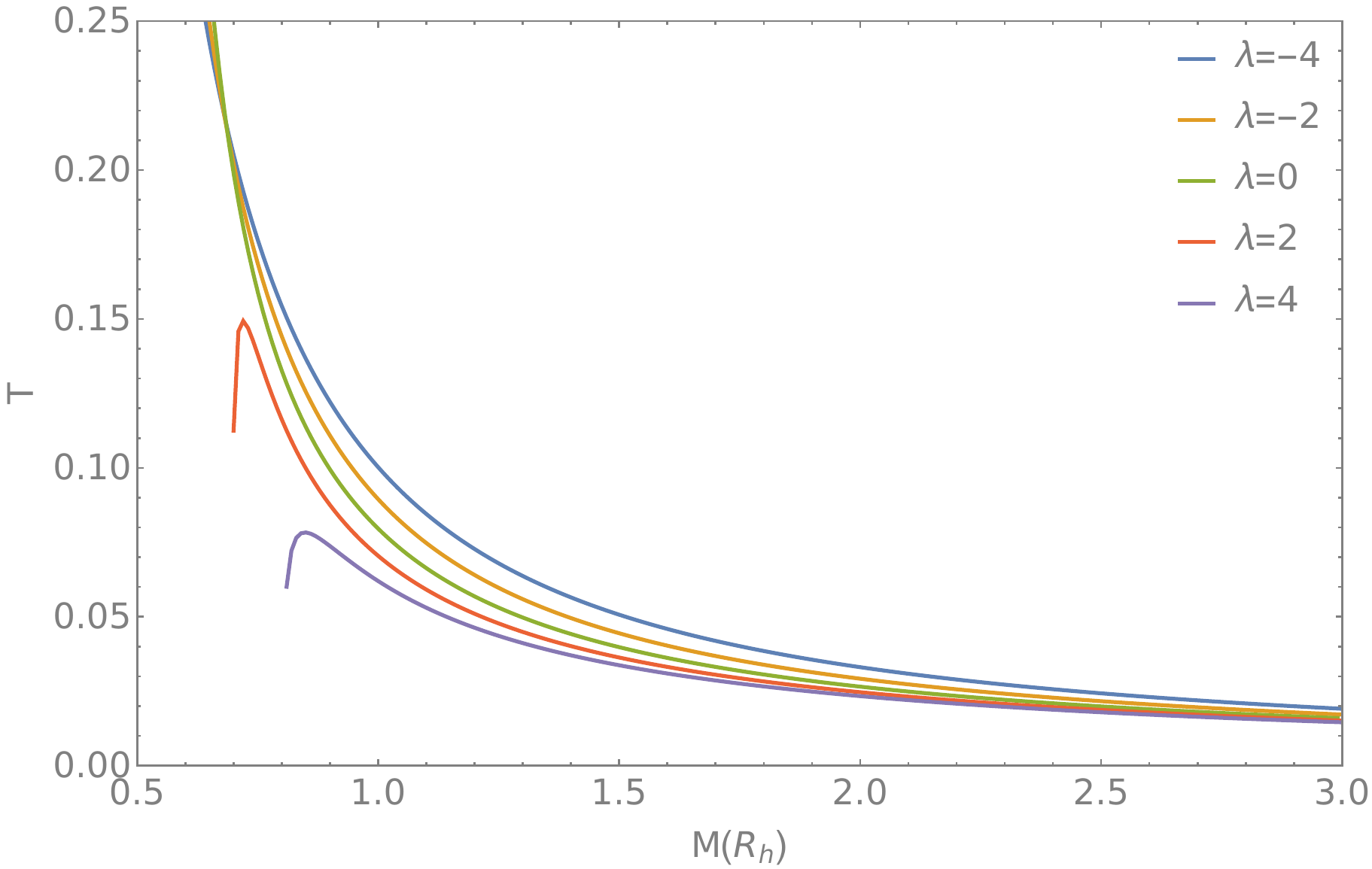}
     \end{subfigure}
        \caption{In this figure we show the energy density $\rho$ as a function of the radial coordinate $r$ (left panel), and the temperature $T$ as a function of the mass of the black hole $M(R_h)$ (right panel). These plots are for the case $\omega=0$ and five different values of the parameter $\lambda$, and assuming $Z=\hbar=1$. }
         \label{fig3}
\end{figure}

Here we analyze the case $\omega = 0$ and, obtain the following solution for $B(r)$
\begin{equation}
    B(r) = 1 - \frac{2 M}{r}+Z r^{-\frac{4 (4 \pi + \lambda)}{16 \pi + \lambda}}.
    \label{eq27}
\end{equation}
From the above equation we see that the metric function $B(r)$ contains information about the fact that we are in the framework of the $f\left(\mathbb{T},\CMcal{T}\right)$ gravity. As expected, the energy density is also modified, and it is given by
\begin{equation}
    \rho(r) = \frac{6 \lambda  Z r^{-\frac{6 (8 \pi + \lambda)}{16 \pi + \lambda}}}{(16 \pi + \lambda)^2}.
    \label{eq28}
\end{equation}
The left panel of Fig. \ref{fig3} shows how the energy density behaves with $r$ and $\lambda$ for the $\omega=0$ case. For the GR case $(\lambda=0)$ there is no field surrounding the black hole since we have $\rho(r)=0$. For positive values of $\lambda$ the energy density diverges for $r \rightarrow 0$ and tends to zero from above for increasing values of the radial coordinate. When we consider negative values of $\lambda$, we obtain $\rho \rightarrow -\infty$ when $r$ goes to zero, and $\rho$ tends to zero from below when $r$ increases. For a fixed value of $r$, $\rho$ increases with the increasing of $\lambda$. The temperature in the present case is as follows
\begin{equation}
    T = \frac{\hbar }{4 \pi  R_h}-\frac{3 \lambda  Z \hbar  R_h^{-\frac{32 \pi + 5 \lambda}{16 \pi + \lambda}}}{4 \pi  (16 \pi + \lambda)}.
    \label{eq29}
\end{equation}
From the right panel of Fig. \ref{fig3} we can observe that when $M(R_h)$ goes towards zero, the temperature, for positive values of $\lambda$, increases until it reaches a value of mass and then it starts to decrease and, for negative values of $\lambda$ and for $\lambda=0$, $T$ increases. When $M(R_h)$ increases, $T$ goes to zero and, in general, $T$ increases with the decreasing of $\lambda$. We notice that for the cases: $\lambda=0$ and $M(R_h)  \lesssim 0.5$; $\lambda=2$ and $M(R_h)  \lesssim 0.7$; and  $\lambda=4$ and $M(R_h)  \lesssim 0.81$, we obtain black hole solutions without a horizon. 

\subsection{Black hole surrounded by a cosmological constant field}

\begin{figure}[t]
     \centering
     \begin{subfigure}{0.49\textwidth}
         \centering
         \includegraphics[width=\textwidth]{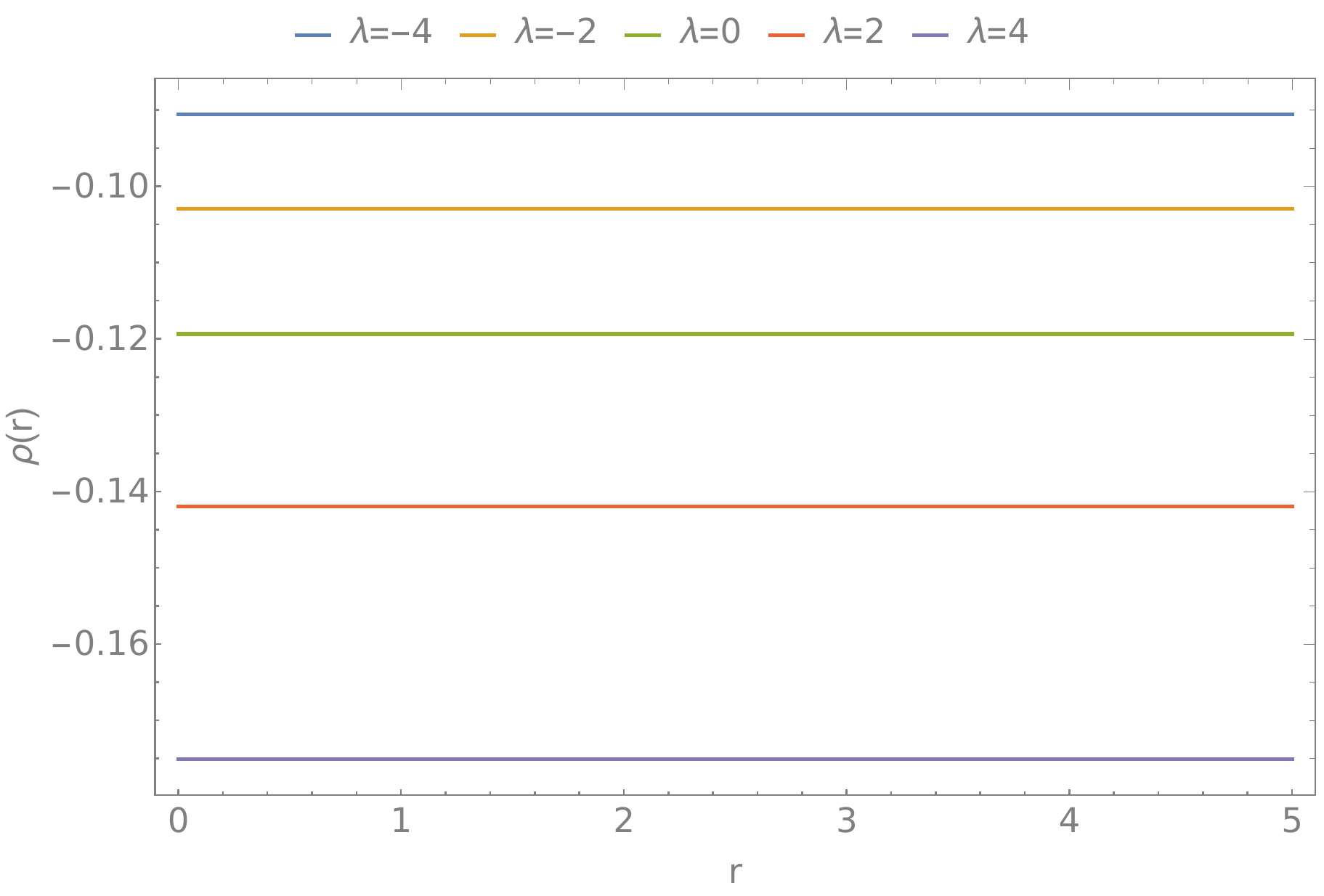}
     \end{subfigure}
     \begin{subfigure}{0.49\textwidth}
         \centering
         \includegraphics[width=\textwidth]{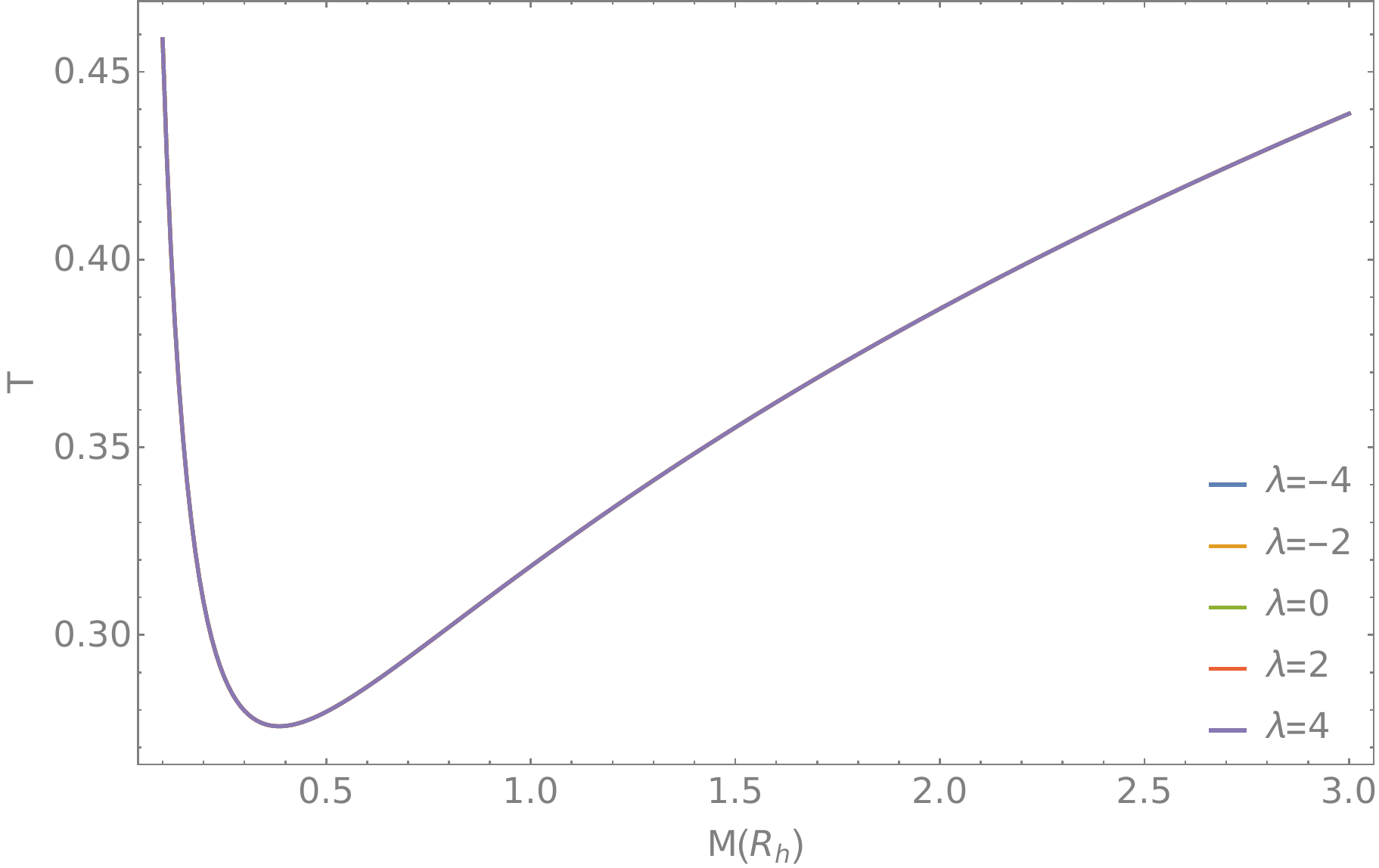}
     \end{subfigure}
        \caption{In this figure we show the energy density $\rho$ as a function of the radial coordinate $r$ (left panel), and the temperature $T$ as a function of the mass of the black hole $M(R_h)$ (right panel). These plots are for the case $\omega=-1$ and five different values of the parameter $\lambda$, and assuming $Z=\hbar=1$. }
         \label{fig4}
\end{figure}

Here we analyze the special case $\omega=-1$. For this choice of $\omega$, $B(r)$ is given by
\begin{equation}
    B(r) = 1 - \frac{2 M}{r}+Z r^2.
    \label{eq30}
\end{equation}
In Eq. (\ref{eq30}) we can observe that in this case the solution for the metric is the same as the one obtained by Kiselev in GR \cite{kiselev}. In other words, the result obtained in the framework of the $f\left(\mathbb{T},\CMcal{T}\right)$ gravity is analogous to the one obtained by Kiselev in GR \cite{kiselev}. This behavior was also observed in Rastall gravity \cite{heydarzade2017black} and in $f(R,T)$ \cite{santos2023kiselev}. Despite having the same metric solution as in GR, the energy density is modified according to
\begin{equation}
    \rho(r) = -\frac{3 Z}{8 \pi - 2 \lambda}.
    \label{eq31}
\end{equation}
We can conclude from the above equation that the energy density is inversely proportional to the $f\left(\mathbb{T},\CMcal{T}\right)$ coupling constant $\lambda$, but it does not depend on the radial coordinate, as can be seen on the left panel of Fig. \ref{fig4}. The Hawking temperature for this case is as follows
\begin{equation}
    T = \frac{\hbar }{4 \pi  R_h}+\frac{3 Z \hbar  R_h}{4 \pi}.
    \label{eq32}
\end{equation}
It is interesting to notice that $T$ also does not depend on the modified theory of gravity taken into account, in which case the Hawking temperature only varies with $M(R_h)$. As shown in the right panel of Fig. \ref{fig4}, $T$ will diverge for $M(R_h) \rightarrow 0$, then for increasing values of $M(R_h)$ the temperature decreases until it reaches a minimum, and then it starts to increase again with the increasing of the mass of the black hole.

\subsection{Black hole surrounded by a quintessence field}

\begin{figure}[t]
     \centering
     \begin{subfigure}{0.49\textwidth}
         \centering
         \includegraphics[width=\textwidth]{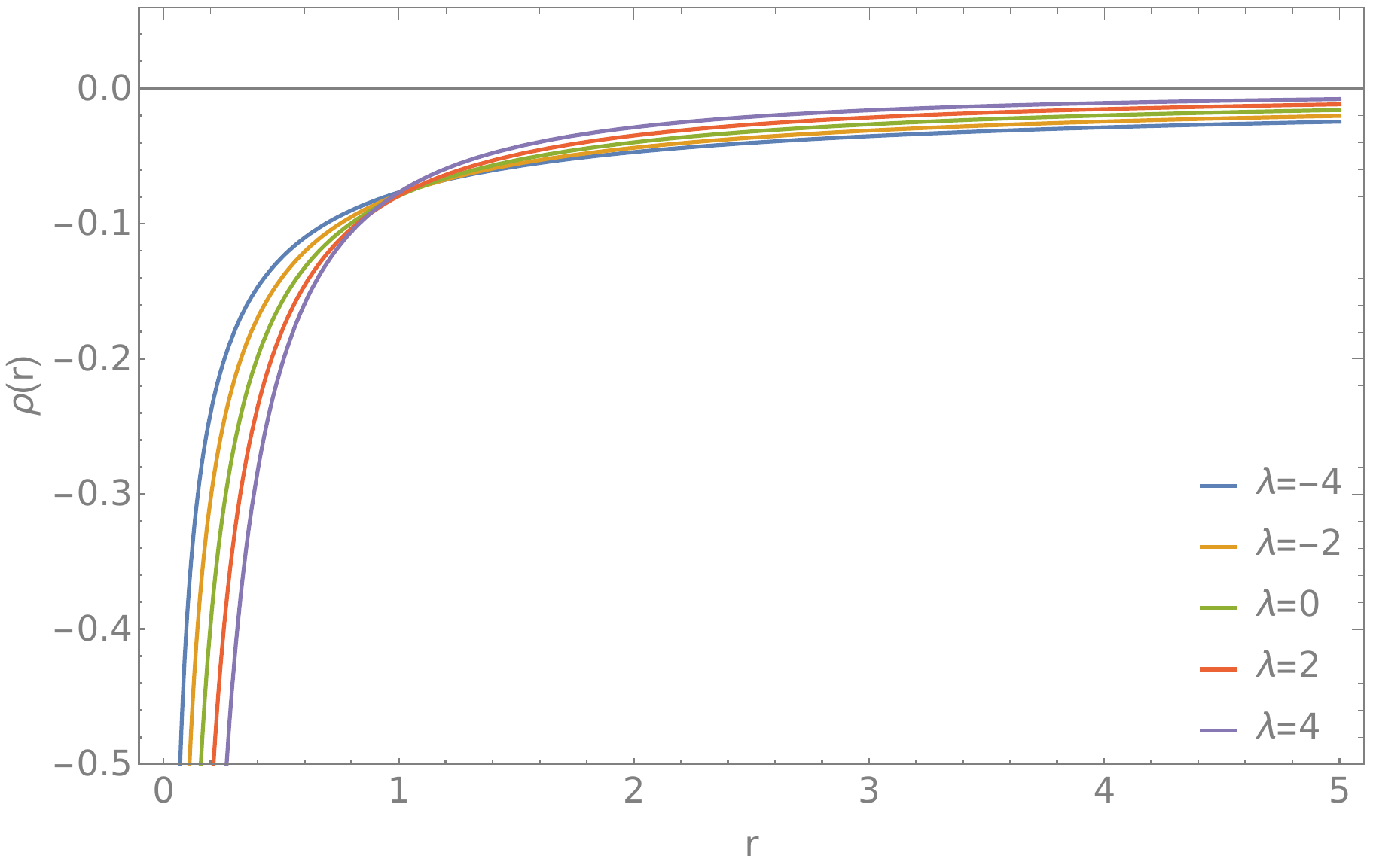}
     \end{subfigure}
     \begin{subfigure}{0.49\textwidth}
         \centering
         \includegraphics[width=\textwidth]{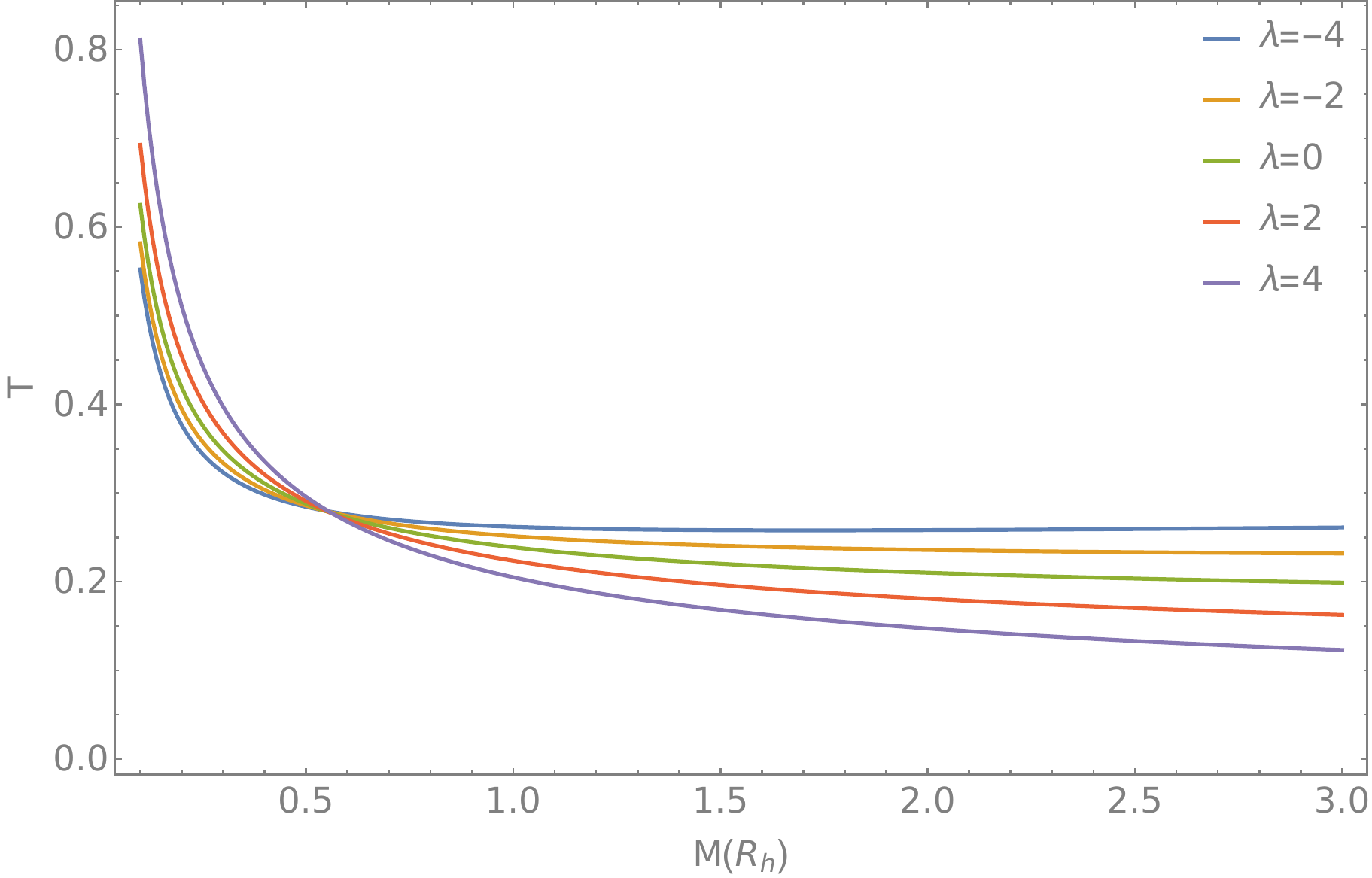}
     \end{subfigure}
        \caption{In this figure we show the energy density $\rho$ as a function of the radial coordinate $r$ (left panel), and the temperature $T$ as a function of the mass of the black hole $M(R_h)$ (right panel). These plots are for the case $\omega=-2/3$ and five different values of the parameter $\lambda$, and assuming $Z=\hbar=1$. }
         \label{fig5}
\end{figure}

Let us now consider the special case $\omega=-2/3$. Thus, the metric function $B(r)$ is
\begin{equation}
    B(r) = 1 - \frac{2 M}{r}+Z r^{-\frac{48 \pi - 20 \lambda }{48 \pi - 7 \lambda }}.
    \label{eq33}
\end{equation}
We can conclude from the above equation that also in this case the spacetime metric codifies the role played by  the modified theory of gravity. The energy density for this case is as follows
\begin{equation}
    \rho(r) = -\frac{18 (32 \pi -9 \lambda ) Z r^{-\frac{6 (8 \pi  + \lambda)}{48 \pi -7 \lambda }}}{(48 \pi -7 \lambda )^2}.
    \label{eq34}
\end{equation}
In the left panel of Fig. \ref{fig5} we show the energy density as a function of the radial coordinate. We can observe that the energy density does not satisfy the energy conditions for none of the values of $\lambda$ analyzed, which is in accordance with the analysis made in Section \ref{energy}. As for the Hawking temperature, it is given by
\begin{equation}
    T = \frac{\hbar }{4 \pi  R_h}+\frac{3 (32 \pi -9 \lambda ) Z \hbar  R_h^{-\frac{13 \lambda }{48 \pi -7 \lambda }}}{4 \pi  (48 \pi -7 \lambda )}.
    \label{eq35}
\end{equation}
We can observe in the right panel of Fig. \ref{fig5} that for all values of the parameter $\lambda$, $T$ diverges when $M(R_h) \rightarrow 0$, and then decreases with the increasing of $M(R_h)$. Also, for this case, the temperature decreases with the increasing of the parameter of the modified teleparallel gravity.

\subsection{Black hole surrounded by a phantom field}

\begin{figure}[t]
     \centering
     \begin{subfigure}{0.49\textwidth}
         \centering
         \includegraphics[width=\textwidth]{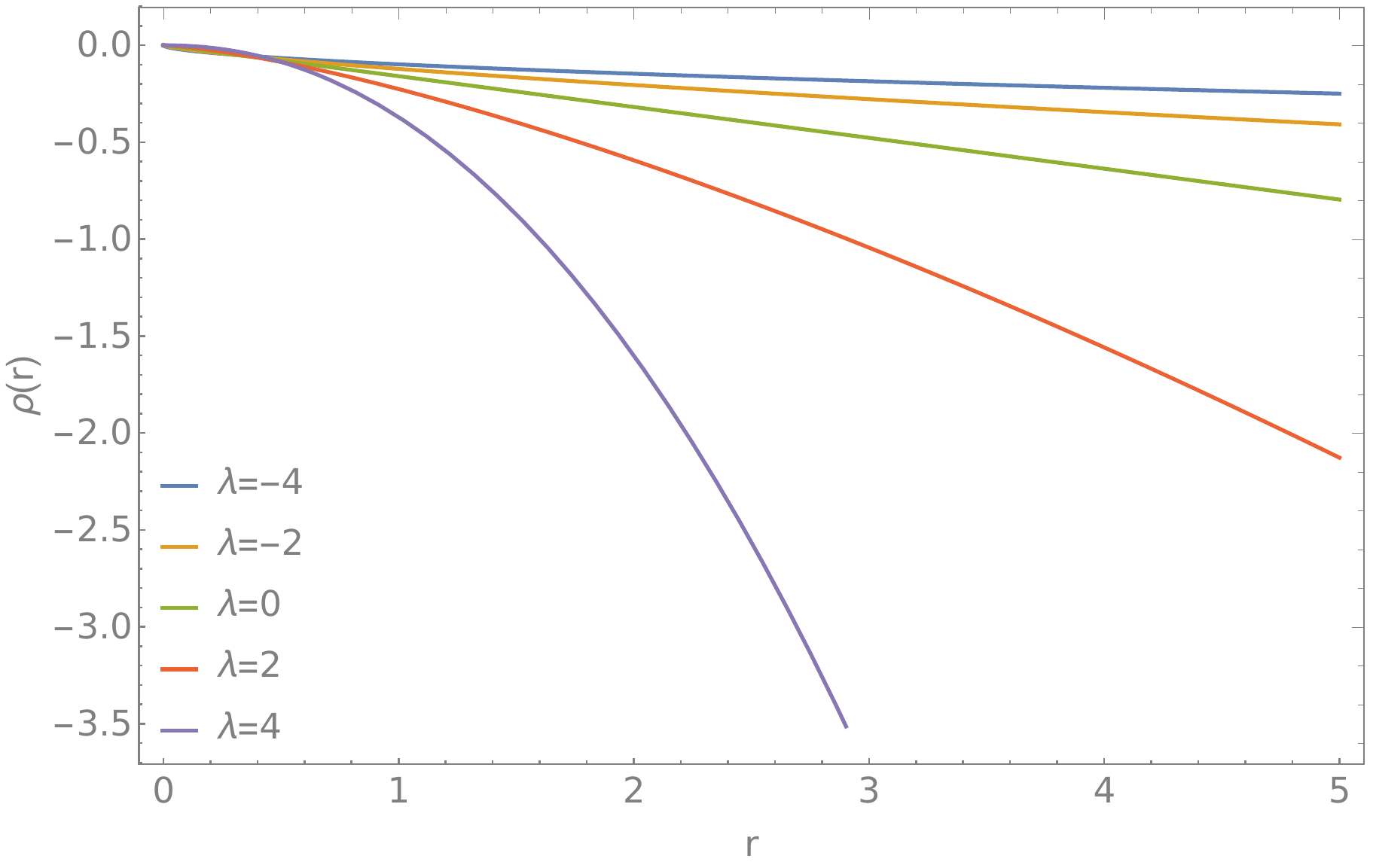}
     \end{subfigure}
     \begin{subfigure}{0.49\textwidth}
         \centering
         \includegraphics[width=\textwidth]{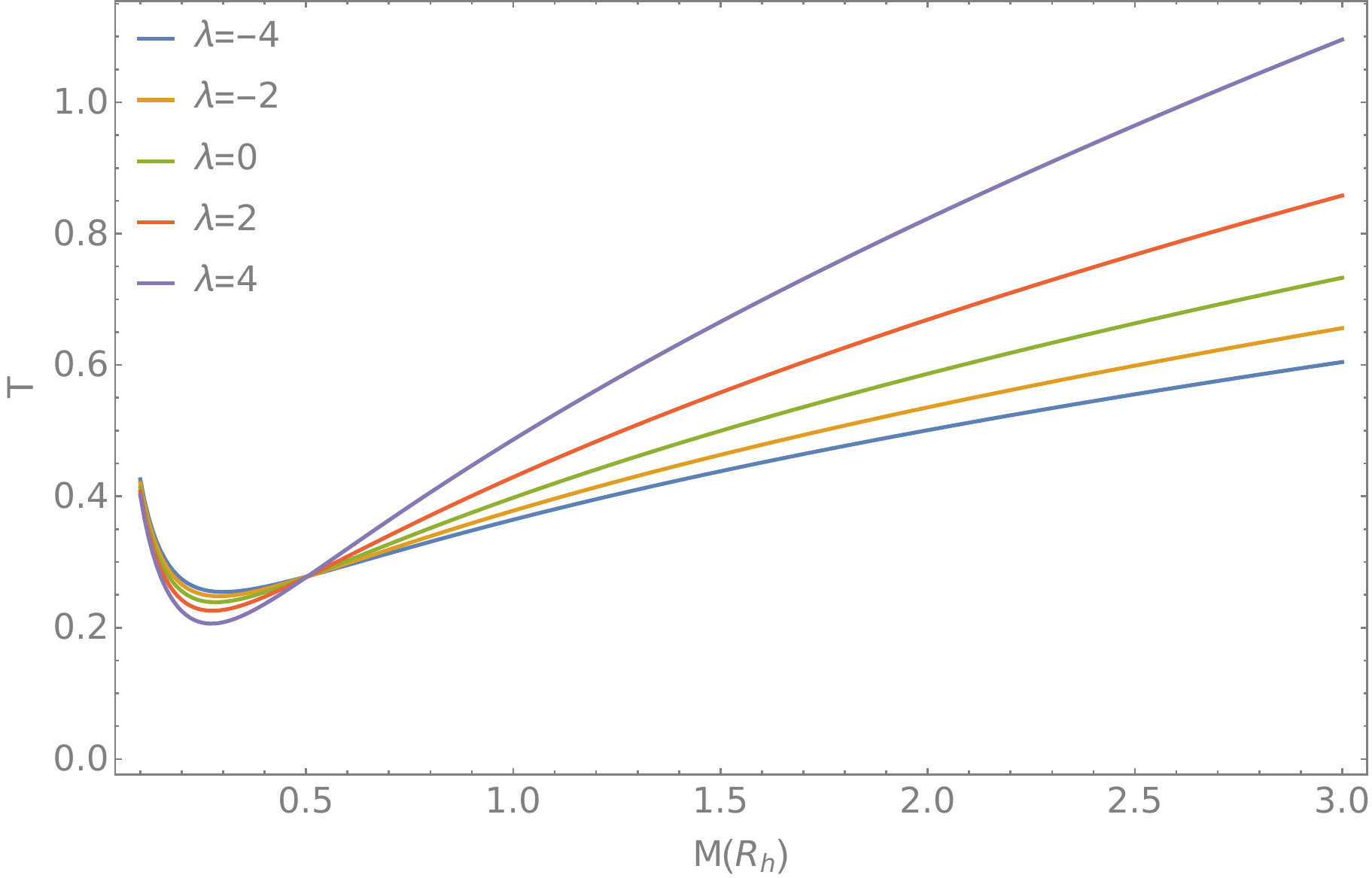}
     \end{subfigure}
        \caption{In this figure we show the energy density $\rho$ as a function of the radial coordinate $r$ (left panel), and the temperature $T$ as a function of the mass of the black hole $M(R_h)$ (right panel). These plots are for the case $\omega=-4/3$ and five different values of the parameter $\lambda$, and assuming $Z=\hbar=1$. }
         \label{fig6}
\end{figure}

Lastly, we study the case $\omega=-4/3$. Now, the metric function $B(r)$ is given by the following expression
\begin{equation}
    B(r) = 1 - \frac{2 M}{r}+Z r^{\frac{4 (36 \pi -7 \lambda )}{48 \pi -17 \lambda }}.
    \label{eq36}
\end{equation}
From the above equation, we conclude that for $\omega=-4/3$, the metric is also affected by the fact that we are in the framework of a modified theory. As for the energy density, it is given by
\begin{equation}
    \rho(r) = -\frac{18 (64 \pi -15 \lambda ) Z r^{\frac{6 (8 \pi + \lambda )}{48 \pi -17 \lambda }}}{(48 \pi -17 \lambda )^2}.
    \label{eq37}
\end{equation}
As in the previous subsection, in this case, we also can observe from the left panel of Fig. \ref{fig6} that in any of the cases analyzed $\rho(r)$ satisfies the energy conditions. This was also expected from the analysis made in Section \ref{energy}. The temperature for the value of $\omega$ we are considering now, is given by
\begin{equation}
    T = \frac{\hbar }{4 \pi  R_h}+\frac{3 (64 \pi -15 \lambda ) Z \hbar  R_h^{\frac{96 \pi -11 \lambda }{48 \pi -17 \lambda }}}{4 \pi  (48 \pi -17 \lambda )}.
    \label{eq38}
\end{equation}
In the right panel of Fig. \ref{fig6}, we can see that $T$ increases when $M(R_h)$ goes to zero, then as the mass of the black increases the temperature decreases until it reaches a minimum, and then it starts to increase again. For this case, we conclude that the temperature increases with the increasing of the parameter $\lambda$ for $M(R_h)  \gtrsim 0.5$.

\section{Final Remarks} \label{conclusions}

In this work we obtained a black hole solution, in the framework of $f(\mathbb{T},\CMcal{T})$ gravity, which represents an extension of the Kiselev black hole to the framework of this modified theory of gravity, and given by Eqs. (\ref{eq15}) and (\ref{eq16}). Our strategy was to identify the fluid under consideration as an anisotropic fluid and to map the components of the Kiselev energy--momentum tensor into the corresponding energy--momentum tensor of the anisotropic fluid. This approach allows us to identify the Lagrangian associated with the fluid and consequently write the field equations for this physical system. We have assumed the function $f(\mathbb{T},\CMcal{T}) = \lambda \mathbb{T} \CMcal{T}$, where $\lambda$ is a coupling constant of geometry with matter fields. This particular form of $f(\mathbb{T},\CMcal{T})$ allows us to obtain an exact solution for the field equations on this theory. The general solution given by Eq. (\ref{eq15}) carries dependence on the coupling constant and on the parameter of the equation of state. Thus, the presence of a modified gravity scenario adds additional terms to the line element that represents the geometry of the spacetime associated to an extension of the Kiselev black hole, in the framework of $f(\mathbb{T},\CMcal{T})$ gravity. 

For this class of solution of the modified field equations, it is possible to explicitly determine the form of the energy density and consequently the radial and tangential pressures by using the equation of state. As we can see in Eq. (\ref{eq16}), the energy density has a general expression in which its behavior can vary greatly, depending on the choice of constituent parameters. In this way, we separately analyze certain values for the parameter of the equation of state. In the same way, we study the general form for the Hawking temperature and analyze some particular values for the parameter of the equation of state. In general, for all special cases analyzed the Hawking temperature goes in the opposite direction than the energy density in relation to the $f(\mathbb{T},\CMcal{T})$ parameter. That is, when $\rho$ increases with $\lambda$, $T$ decreases, and vice-versa, except for the case $\omega=-1$, where the temperature does not varies with $\lambda$. We also concluded that the energy conditions were only satisfied for the cases $\omega=1/3$ and $\omega=0$. As also observed in other modified theories \cite{heydarzade2017black,santos2023kiselev}, in the case  $\omega=-1$ the spacetime metric is not changed by the modified theory of gravity.

The solution obtained in the present work can be used in future works to study other physical quantities such as geodesics associated with light path and massive bodies in addition to the thermodynamics of this black hole considering criticality and efficiency. The rotating black hole solution surrounded by a Kiselev fluid in $f(\mathbb{T},\CMcal{T})$ gravity is another example of future work based on our results.  

\section*{Acknowledgements}
L.C.N.S would like to thank FAPESC for financial support under grant 735/2024.

\bibliographystyle{ieeetr}
\bibliography{ref.bib}

\end{document}